\documentclass[a4paper,fleqn,usenatbib]{mnras}

\usepackage{natbib}

\usepackage[T1]{fontenc}
\usepackage{ae,aecompl}

\usepackage{graphicx}	
\usepackage{amsmath}	
\usepackage{amssymb}	

\title[Energy-dependent Time lag and  rms]{A model for the energy-dependent time-lag and rms of the heartbeat oscillations 
in GRS 1915+105}
\author[M.H.Mir et al.]{Mubashir Hamid Mir,$^{1,2}$\thanks{E-mail: mubiphst@gmail.com}
Ranjeev Misra,$^{2}$
Mayukh Pahari,$^{2}$ \newauthor Naseer Iqbal$^{1}$ and Naveel Ahmad$^{1}$
\\
$^{1}$ Department Of Physics, University of Kashmir, Srinagar-190006, India \\
$^{2}$ Inter-University Center for Astronomy and Astrophysics, Post Bag 4, Ganeshkhind, Pune-411007, India}

\begin{document}

\label{firstpage}
\pagerange{\pageref{firstpage}--\pageref{lastpage}}
\maketitle

\date{}

\label{firstpage}

\begin{abstract}
Energy dependent phase lags reveal crucial information about the causal relation
between various spectral components and about the nature of the
accretion geometry around the compact objects.
The time-lag and the 
 fractional root mean square (rms) spectra of GRS 1915+105 in its heartbeat oscillation
 class/$\rho$ state show peculiar behaviour at the fundamental and
 harmonic frequencies where the lags at the fundamental show a
 turn around at $\sim$ 10 keV while the lags at the harmonic do not show any turn around at least till $\sim$ 20 keV.  The magnitude of lags are of the order
 of few seconds and hence cannot be attributed to the light travel time
 effects or Comptonization delays. The 
 continuum X-ray spectra can roughly be described by a disk blackbody and a
 hard X-ray power-law component and from phase 
resolved spectroscopy it has been shown that the inner disk radius varies during the
oscillation. Here, we propose that there is a delayed response of the inner disk radius (DROID) to the accretion rate such that 
 $r_{in}(t)\propto \dot{m}^\beta (t-\tau_d)$. The fluctuating
 accretion rate drives the oscillations of the inner radius after a
 time delay $\tau_d$ while the power-law component responds immediately.
We show that in such a scenario a pure sinusoidal oscillation of the accretion rate
can  explain not only the shape and magnitude of energy dependent rms and time-lag spectra at the fundamental
but also the next harmonic with just four free parameters. 
\end{abstract}

\begin{keywords}
accretion, accretion discs$-$black hole physics$-$X-rays: binaries$-$X-rays: individual: GRS 1915+105
\end{keywords}

\section{Introduction}
   
GRS 1915+105 exhibits a range of variability in its light curve and
power density spectra (PDS), along with other peculiar features like
near-Eddington accretion rate consistent for 20 years, jet activities
$-$ from steady radio flickering to super-luminal radio ejections
\citep{Mi94} and regular transitions into various spectral states in
time-scale from msec to hours \citep{MR97,MM99,Mi94,BM197}.
The timing analysis features 14 variability classes with each having
their own peculiar light curves, phase/time-lags and coherence
\citep{BM197,pa13a,pa13b}. Among the 14 variability classes, the $\rho$
class shows characteristic limit-cycle variability in the time-scale
of 50$-$100 sec with large amplitude oscillations in X$-$ray intensity
by nearly an order of magnitude \citep{TC97,Bel97}.  
 
A promising model for the  $\rho$ class variability is that it is
driven by radiation pressure instability \citep{Le74,Tl84,TC97}.
However, numerical simulations of the instability suggests that there
is need to change the viscous prescription and take into account
a fraction of energy dissipating in a corona to explain the overall
X-ray modulation \citep[e.g.][]{Nay00,CZ2000,CZ2005}.  Phase resolved spectroscopy
of the variability reveals a consistent picture where the instability
causes the inner disk radius to vary with the luminosity variation,
 which may be due to mass ejection from the system \citep{Ne11,Ne12}. The
spectral modeling at different phases show complex components consisting
of disk emission and a power-law with a high energy cutoff.

\begin{figure*}
 \centering
  \includegraphics[width=0.34\textwidth,angle=-90]{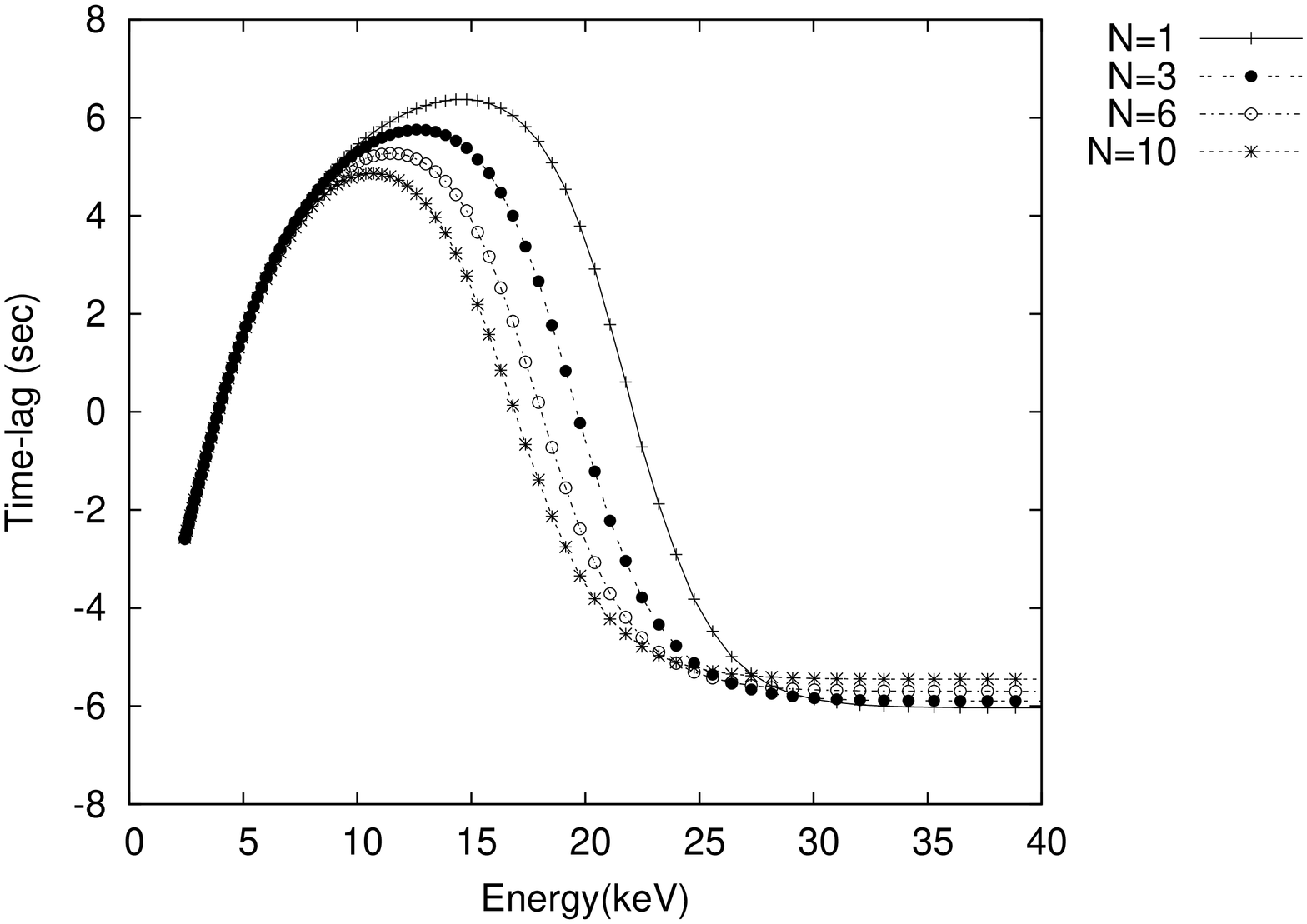}
  \includegraphics[width=0.34\textwidth,angle=-90]{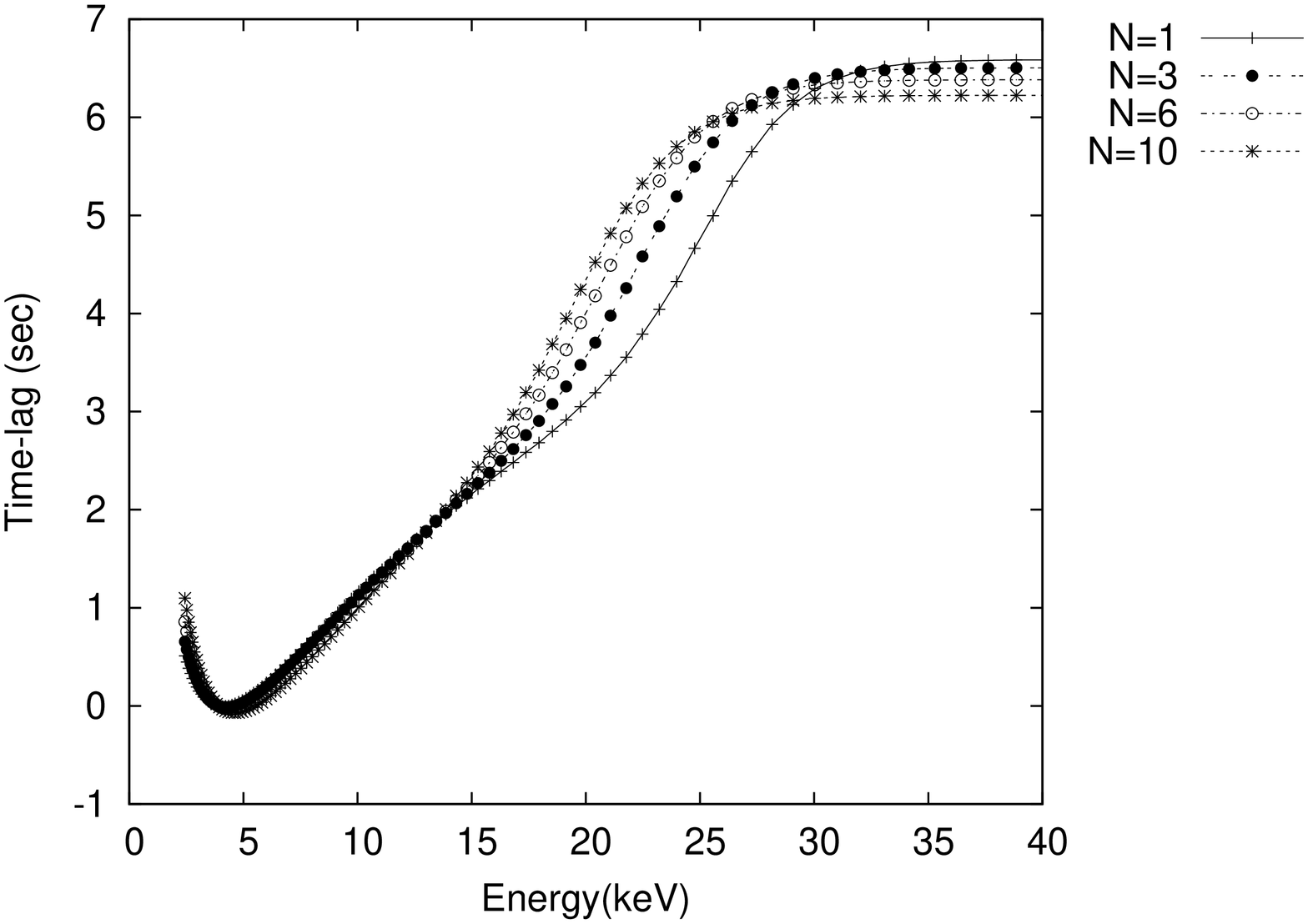}
  \includegraphics[width=0.34\textwidth,angle=-90]{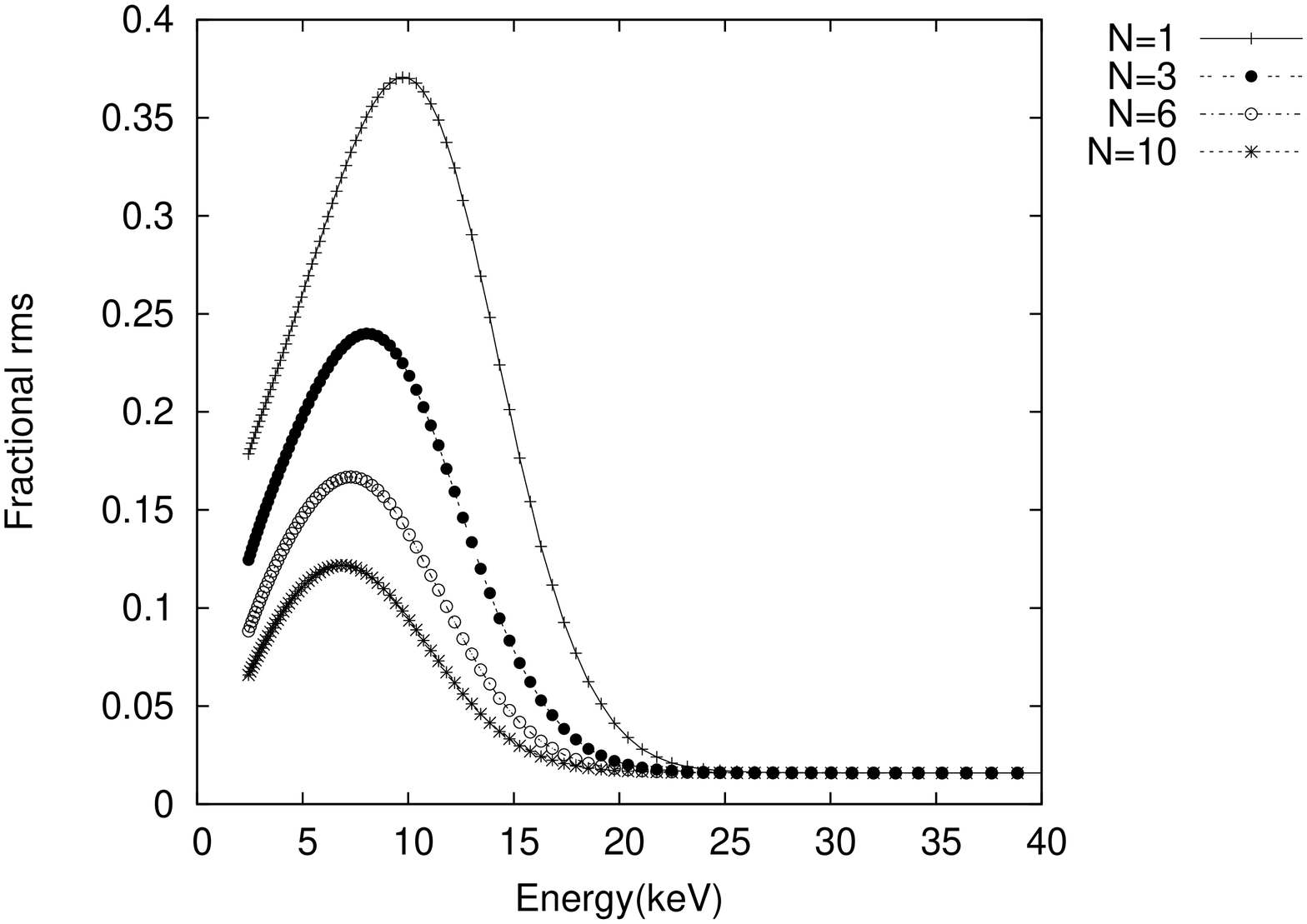}
  \includegraphics[width=0.34\textwidth,angle=-90]{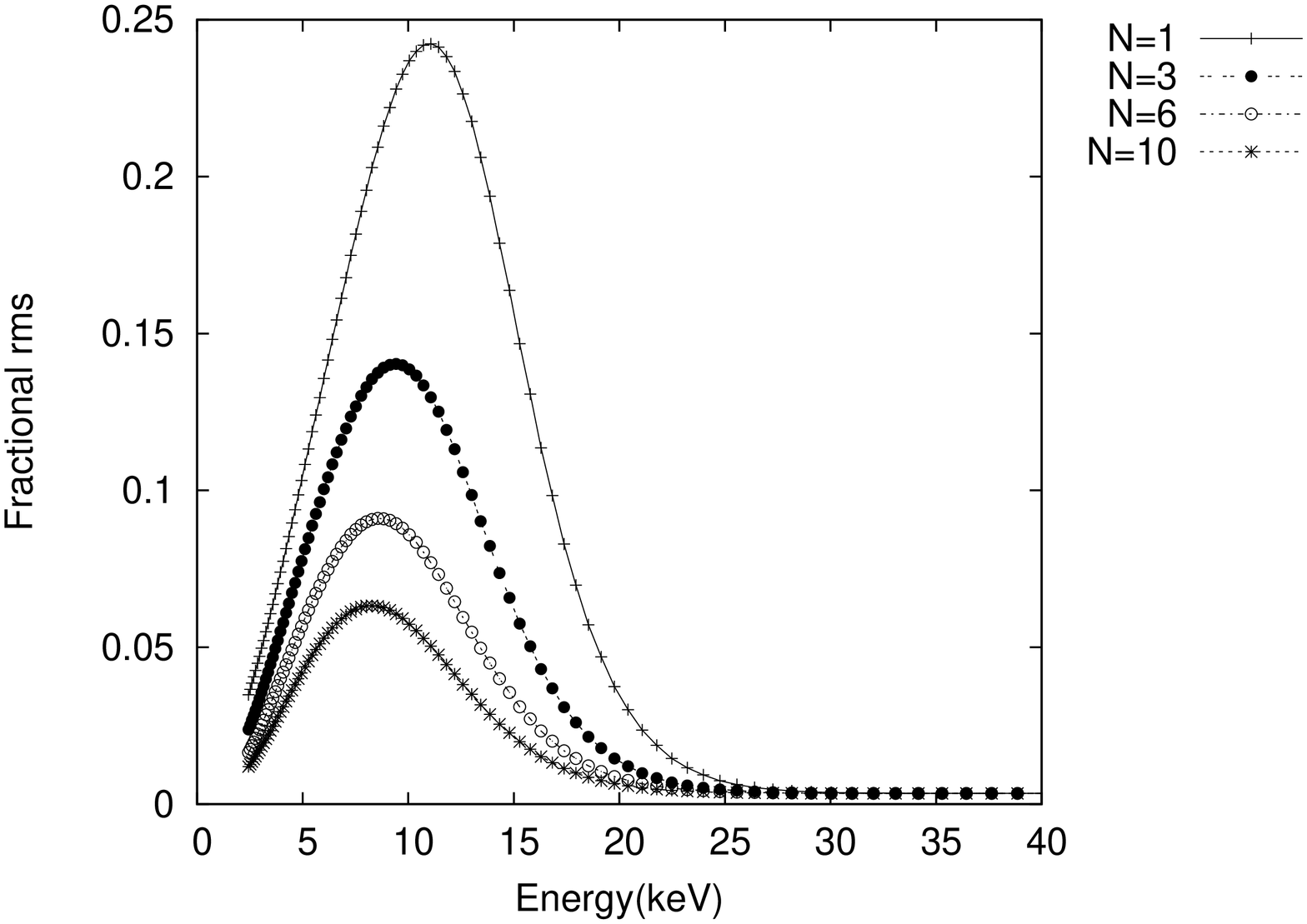}
  \caption{The predicted time-lag and rms versus energy for the fundamental (left panels) and the next harmonic (right panels)
for different values of the power-law to disk flux ratio parameter $N$. The other parameters are kept at some constant fiduciary values.
Note the sensitivity of the peak of the time-lag curve for the fundamental to $N$.}
\label{Nvar}
\end{figure*}

An alternate approach to phase resolved spectroscopy is to understand
the energy dependent phase lag and rms spectra of an oscillation.
This is particularly useful for Quasi periodic oscillations (QPO).  
Such analysis of high frequency QPOs can give crucial information
regarding the size of the emitting region  \citep[e.g.][]{LM01,Kum14}
and for the low frequency ones the causal connection between
spectral parameters \citep{MM13}. \citet{Ing15} have considered both energy dependent rms and phase resolved
spectroscopy to argue that the low frequency QPO in the $\chi$ state has a geometric origin.
 Frequency resolved spectroscopy at the QPO frequency
which is related to the energy dependent rms curves can provide crucial
information regarding the nature of the oscillation \citep{Axe14}.  \citet{CZ2005} report
hard photon time-lags for the $\rho$ class variability which they interpret
as the delay for the corona to adjust to a varying accretion rate. Indeed
the magnitude of the time delays of the order of seconds implies that
it is not due to light travel time effects and instead should be 
associated with  time delays between different structural parameters.
Since the inner disk radius is known to vary during the oscillation
\citep{Ne12}, it would be interesting to see if  energy dependent
time-lags and the rms spectra of the different harmonics of the
oscillation can provide insight into how the the inner radius varies
with the accretion rate.

As we describe below the time-lag and rms spectra for the $\rho$ class
variability is fairly complex and our motivation is to identify the simplest
model which can describe not only the behaviour of the fundamental but also
simultaneously that of the next harmonic.

\section{The Delayed Response of the Inner Disk Model(DROID)}

\begin{figure*}
 \centering
    \includegraphics[width=0.34\textwidth,angle=-90]{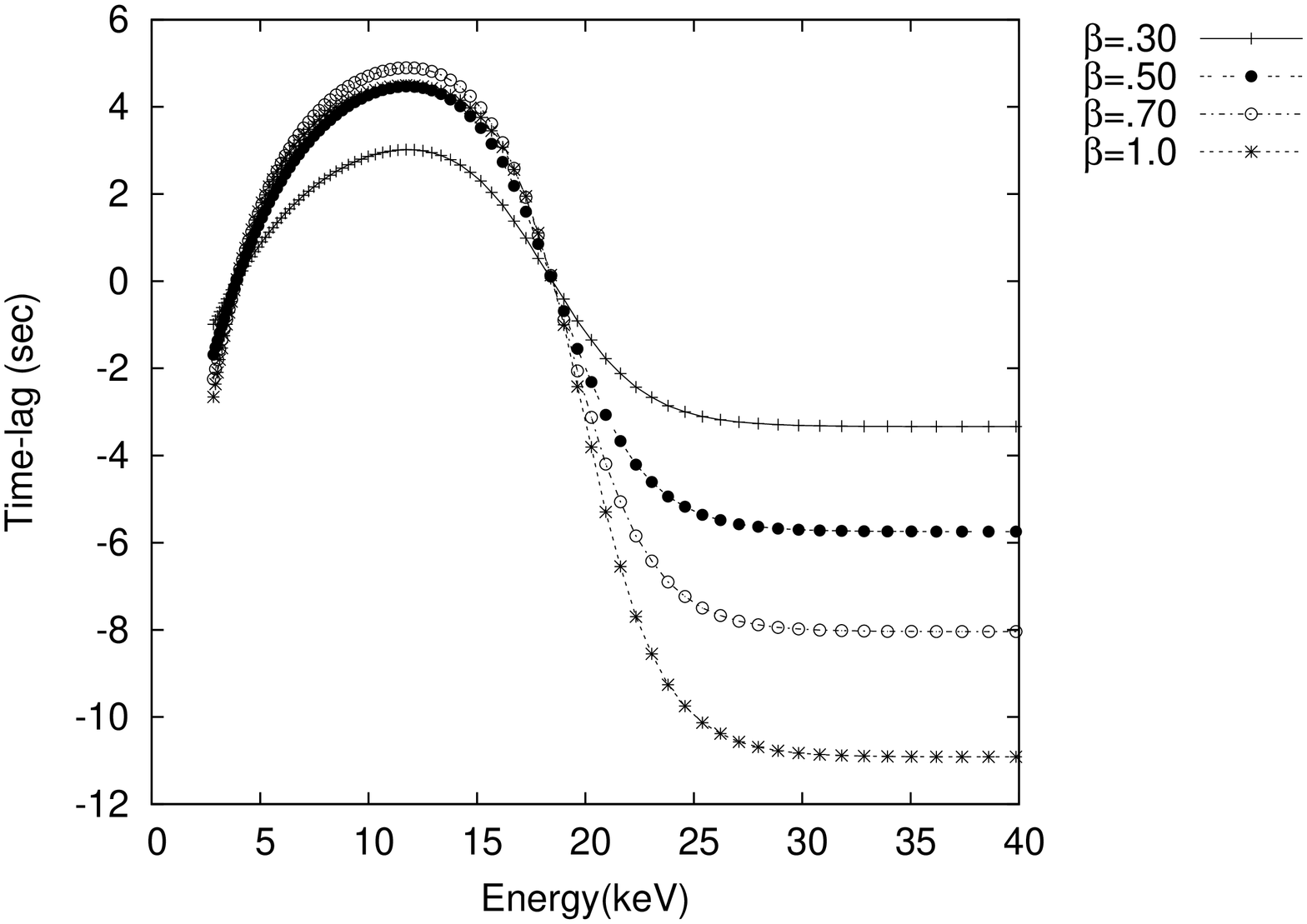}
    \includegraphics[width=0.34\textwidth,angle=-90]{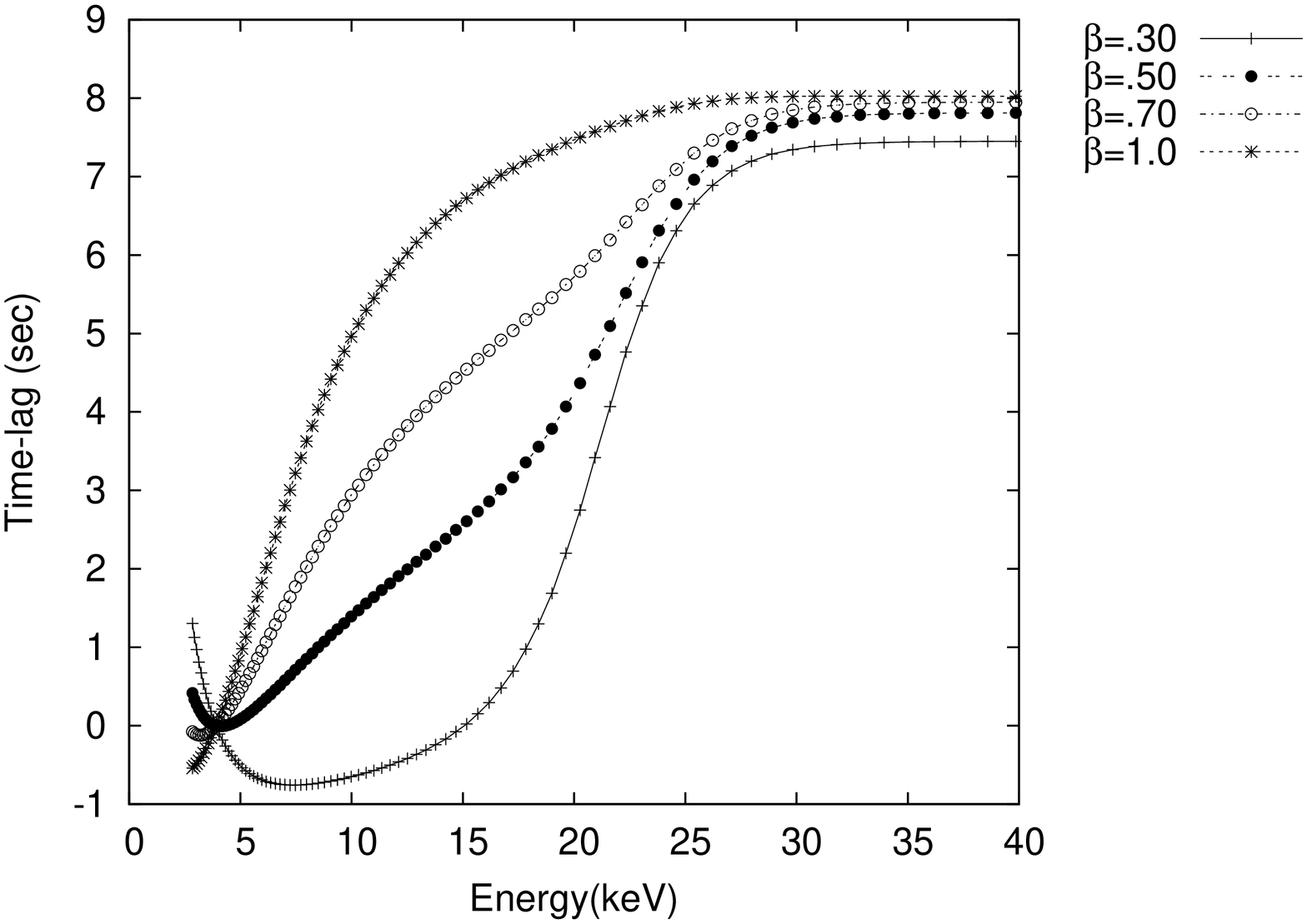}
          \includegraphics[width=0.34\textwidth,angle=-90]{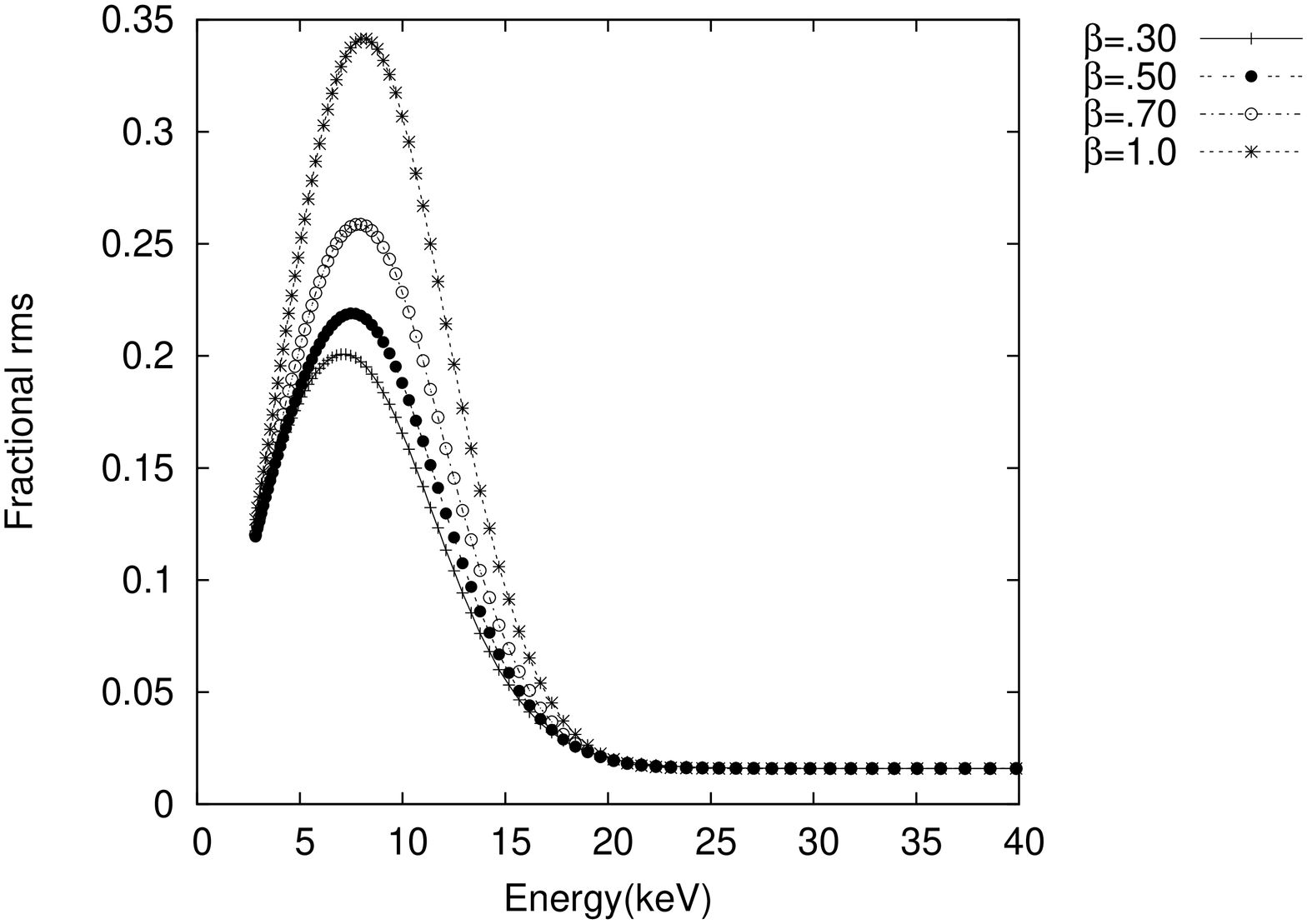}
\includegraphics[width=0.34\textwidth,angle=-90]{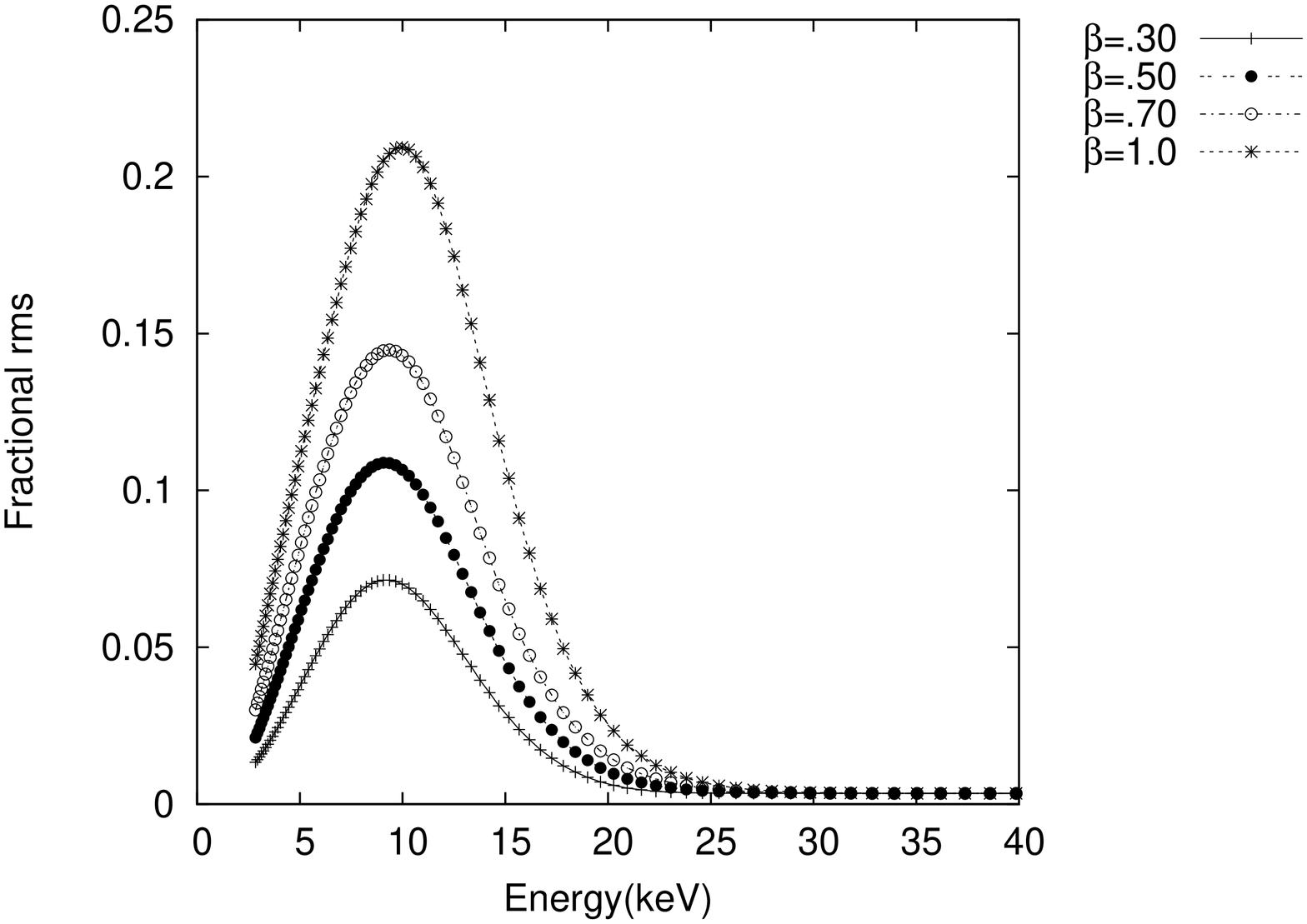}

  \caption{The predicted time-lag and rms versus energy for the fundamental (left panels) and the next harmonic (right panels)
for different values of the parameter $\beta$ which relates the dependence of the inner radius with the accretion rate. 
The other parameters are kept at some constant fiduciary values.
Note the sensitivity of  the time-lag curve for the next harmonic to $\beta$.}
\label{betavar}
\end{figure*}

We assume that the X$-$ray emission 
primarily consists of two components,  a soft disk blackbody
and  a hard X$-$ray power-law extending from $\sim$ 1 keV to high energies
due to a corona. We neglect contributions from the Iron line emission
and other reflection features..

\begin{figure*}
 \centering
    \includegraphics[width=0.3\textwidth,angle=-90]{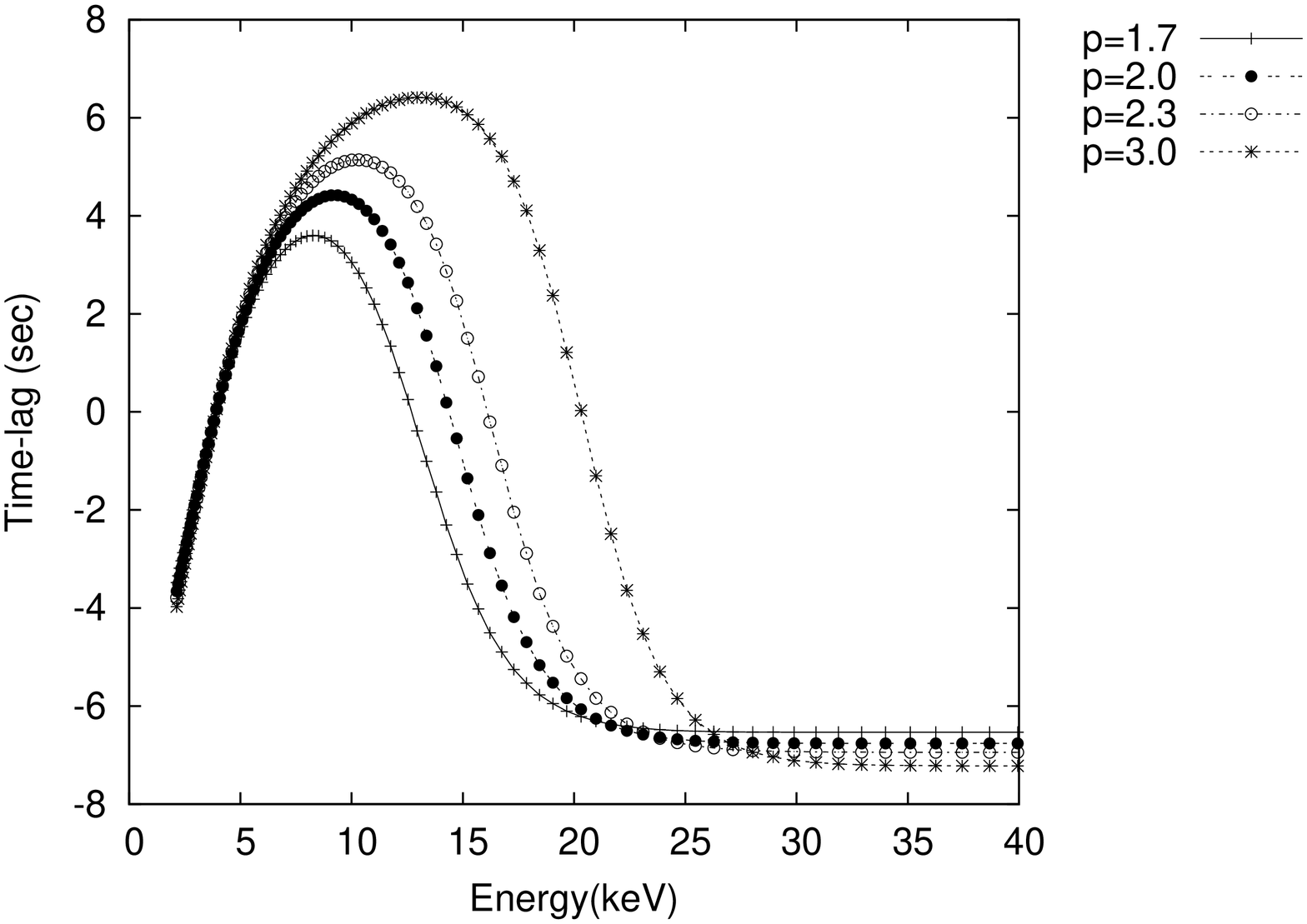}
    \includegraphics[width=0.3\textwidth,angle=-90]{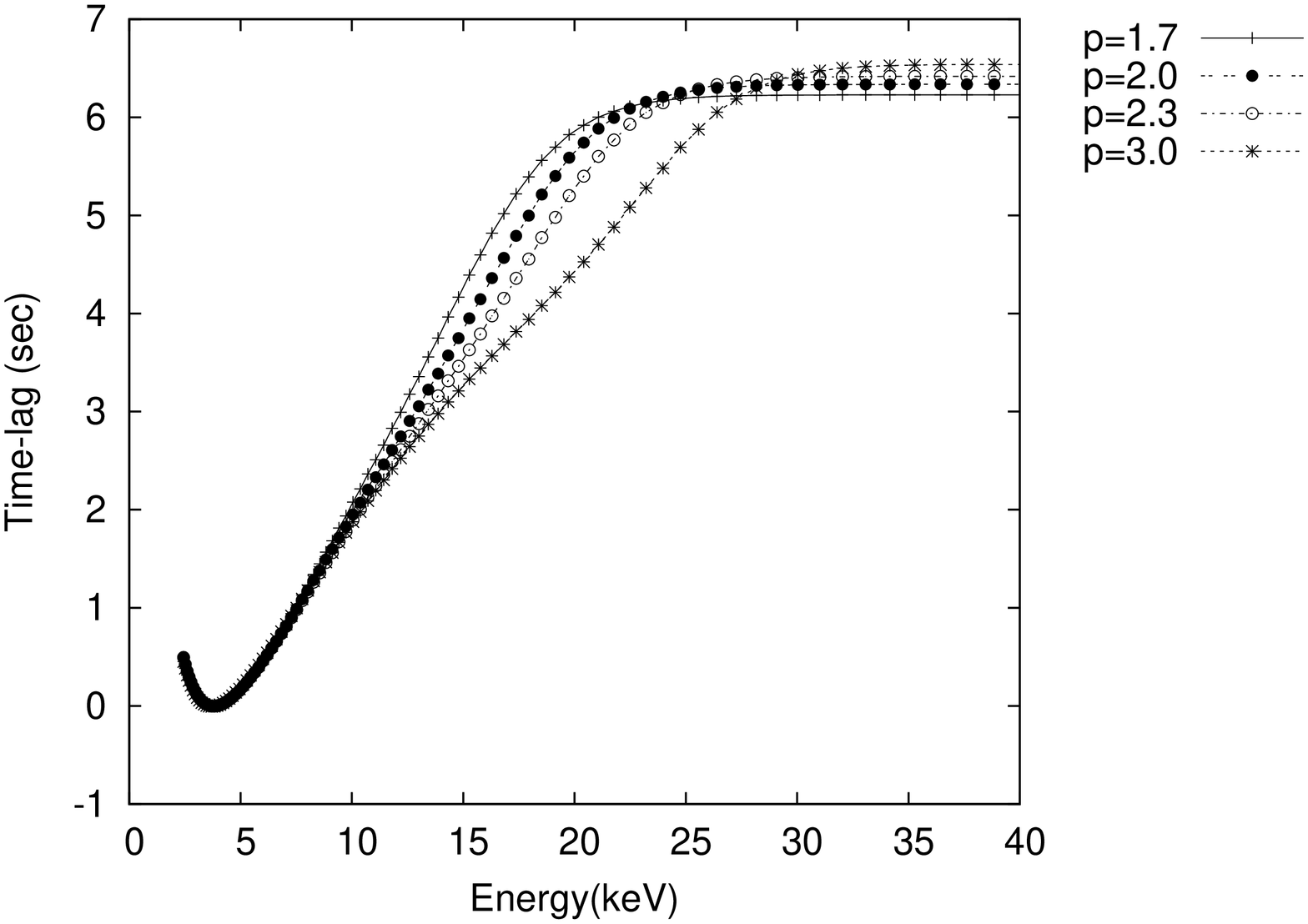}
          \includegraphics[width=0.3\textwidth,angle=-90]{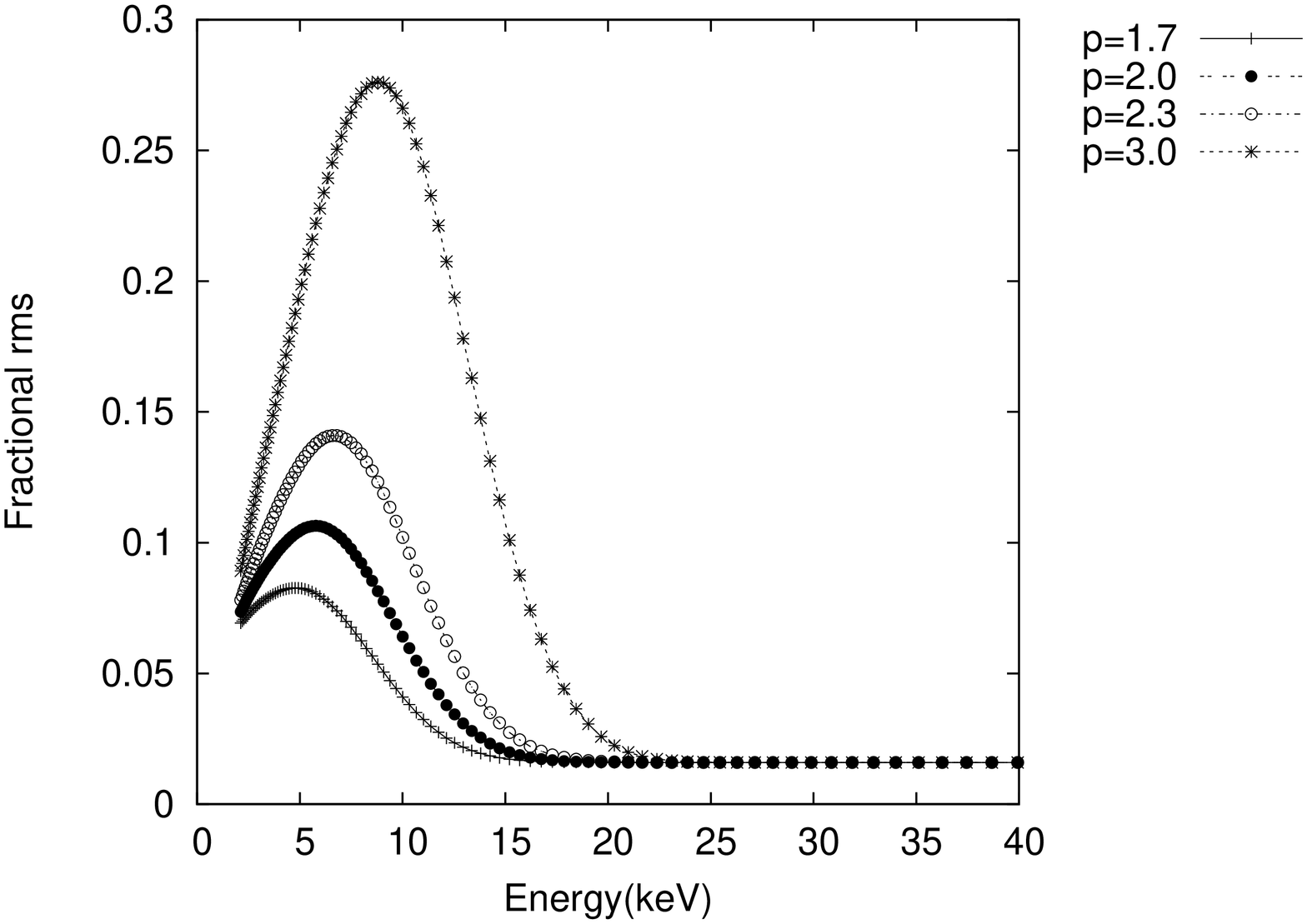}
\includegraphics[width=0.3\textwidth,angle=-90]{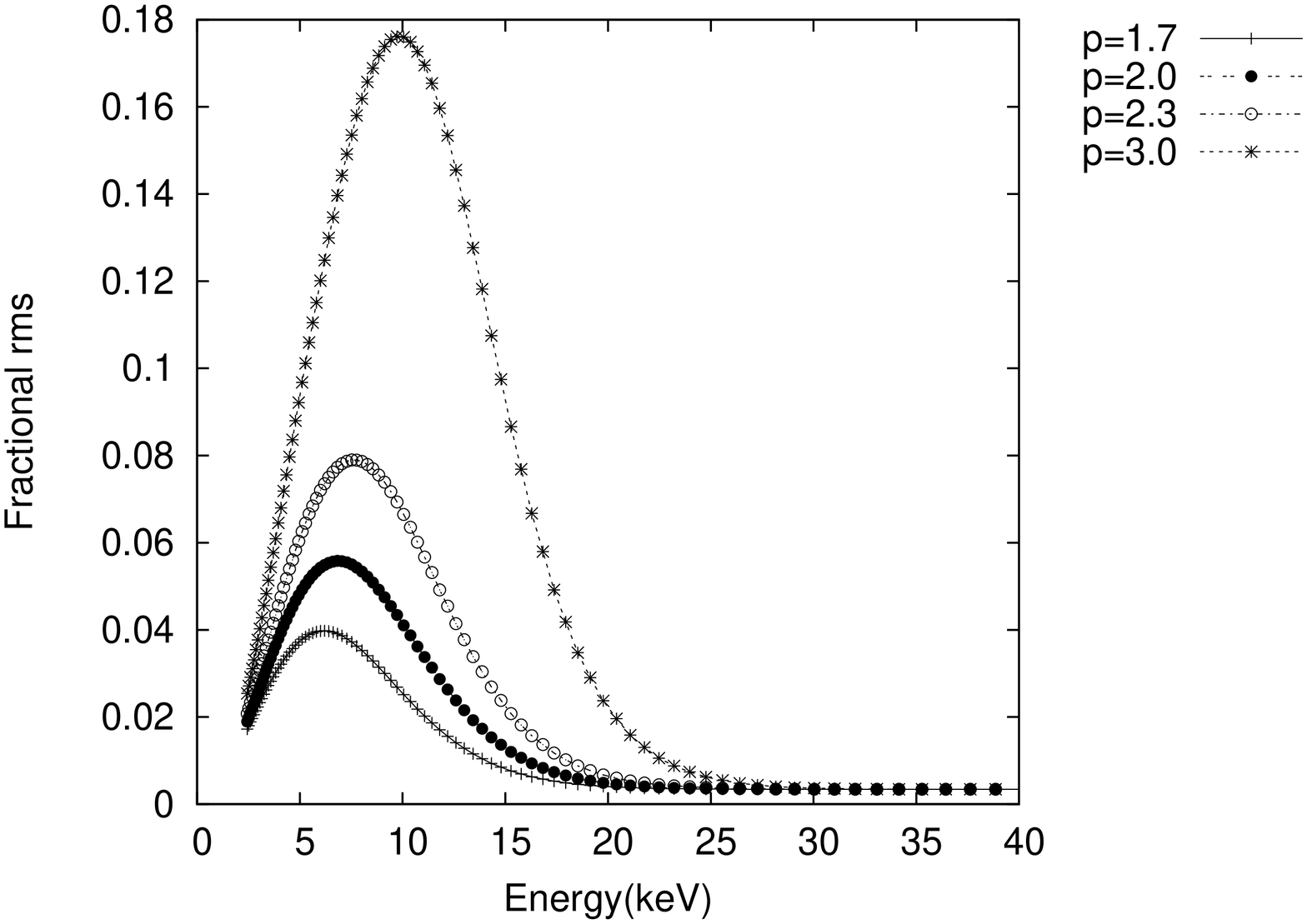}

  \caption{The predicted time-lag and rms versus energy for the fundamental (left panels) and the next harmonic (right panels)
for different values of the power-law index $p$. The other parameters are kept at some constant fiduciary values.}
\label{powvar}
\end{figure*}

The blackbody disk flux is given by
\begin{equation}
F_{d} \propto  E^2 \int_{r_{in}}^{\infty} \frac{r}{exp(\frac{E}{kT(r)})-1}dr
\end{equation}
where $E$ is the energy of the photon, $r$ is the radius, $r_{in}$ is the
inner disk radius
and T(r) is the surface disk temperature. From the standard disk
blackbody model, we have \citep{FKR2002}:
\begin{equation}
kT_{r}=kT_{in}\left(\frac{r}{r_{in}}\right)^{-3/4}
\end{equation}
where $T_{in} \propto \dot{m}^{1/4}{r_{in}}^{-3/4}$, 
$\dot m$ is the accretion rate. The power-law component
is taken to be
\begin{equation}
F_p \propto {\dot m}^\Gamma E^{-P}
\end{equation}
where it has been assumed that the normalization of the power-law 
depends on the accretion rate ($\propto {\dot m}^\Gamma$) while for
simplicity the photon index $P$ does not vary.  

The variation of the total flux $F_T = F_d + F_p$ due to variations
of the accretion rate and inner radius, to second order is given by
\begin{eqnarray}
\delta F_t & = &(F_t)_{\dot{m}}\delta{\dot{m}} + {(F_t)_{r_{in}}} \delta {r_{in}} + \frac{1}{2}(F_t)_{\dot{m}\dot{m}}\delta{\dot{m}^2} \nonumber \\
& & +(F_t)_{\dot{m}r_{in}}\delta{\dot{m}}\delta{r_{in}}+\frac{1}{2}(F_t)_{r_{in}r_{in}} \delta {r_{in}}^2 
\label{eqndft}
\end{eqnarray}
Here we denote the normalized variation 
$\delta X = \Delta X/X$ and define 
\begin{equation}
(Z)_X = \frac{X}{Z}\frac{\partial Z}{\partial X}\;\;\;\; \hbox{and}\;\;\;\;  (Z)_{XY} = \frac{XY}{Z}\frac{\partial }{\partial X}\frac{\partial Z}{\partial Y}
\end{equation}

We assume that the inner disk radius follows the accretion rate with a
time delay defined by:
\begin{equation}
r_{in}(t) \propto \dot{m}^{-\beta}(t-\tau_d)
\end{equation}
and that the accretion rate undergoes a pure sinusoidal variation at an
angular frequency $\omega$
\begin{equation}
\dot{m}(t)= \dot{m_o}(1+\delta\dot{m}) =\dot{m_o}(1+|\delta\dot{m}|e^{i \omega t})
\end{equation}
which drives the variability of the inner radius, $r_{in}(t)  =  r_{ino}(1+\delta r_{in})$
where
\begin{equation}
\delta r_{in} =  -\beta|\delta\dot{m}| e^{i\omega(t-\tau_d)} +\frac{\beta (\beta+1)}{2}|\delta\dot{m}|^2 e^{2i\omega(t-\tau_d)} 
\end{equation}
Substituting the above in Eqn (\ref{eqndft}), and collecting terms proportional to $e^{i\omega t}$ and $e^{2i\omega t}$,
for the fundamental oscillation, we get:
\begin{equation}
\delta F_t^f  = \frac{|\delta \dot{m}|}{1+\frac{F_p}{F_d}}\left(Ae^{i\omega t} + Be^{i\omega (t-\tau_d)}\right)
\label{eqndftf}
\end{equation}
and for the next harmonic:
\begin{equation}
\delta F_t^h=\frac{|\delta\dot{m}|^2}{1+\frac{F_p}{F_d}}\left(A^\prime e^{2i\omega t}+B^\prime e^{2i\omega(t-\tau_d)}+C^\prime e^{2i\omega(t-\tau_d/2)}\right) 
\label{eqndfth}
\end{equation}
The coefficients $A, B, A^\prime, B^\prime$ and $C^\prime$ are derived and listed in Appendix (A).
The ratio of the power-law to disk flux $F_p/F_d$ can be conveniently written as
\begin{equation}
\frac{F_p}{F_d} = N G(E,kT_{in},P)
\end{equation}
The function $G$ is given in Appendix B and $N$ is a parameter which can be obtained
from the time-averaged spectrum. 

The energy dependent rms for the fundamental and next harmonic are given by $\frac{1}{\sqrt{2}}|\delta F_t^f (E)|$
and $\frac{1}{\sqrt{2}}|\delta F_t^h (E)|$ while the phase lag with respect to a reference energy $E_{ref}$ is given by
the phase of $\delta F_t^f (E) (\delta F_t^f (E_{ref}))^*$ and $\delta F_t^h (E) (\delta F_t^h (E_{ref}))^*$.

The energy dependent variations $\delta F_t^f$ and $\delta F_t^h$ given by Equations (\ref{eqndftf}) and (\ref{eqndfth}) depend
on several parameters which can be divided into two groups. The parameters of the first group can be estimated
from the time averaged spectrum which are  the inner disk temperature $T_{in}$, the high energy photon index $P$ and  $N$ 
which parametrizes the ratio of the power-law to disk flux. The parameters of the second group pertain to the nature
of the oscillation and are the exponents for inner disk radius ($\beta$), the power-law flux ($\Gamma$) with
the accretion rate, $\omega t_d$ which is the phase difference (between the inner disk radius \& the accretion rate) and
an over all normalization of the variation $|\delta \dot{m}|$. Thus in this simplistic model, the energy 
dependent rms and time-lag for the fundamental and next harmonic are completely specified by these four
parameters. Of these the normalization of the variability $|\delta \dot{m}|$ determines the normalization of
the rms of the fundamental while its square determines that of the next harmonic.

To illustrate the dependence of the predicted energy dependent rms and lag of this model on parameters,
we show as an example in Figure (\ref{Nvar}) the predicted curves for different values of the
power-law to disk flux ratio characterized by $N$ while the other parameters of the model
are fixed at some fiduciary values i.e.   $T_{in}$, $\beta$,
$p$, $\omega*\tau_d$, $\Gamma$ \& $\Delta\dot{m}$ at $1.5$ keV, $0.65$, $2.7$, $2.2$
,$0.05$ and $0.4$ respectively. A salient feature of the model is that the time-lag
increases with energy and then decreases. This occurs because the high energy photons are
dominated by the power-law component which varies with the accretion rate without any time-delay.
On the other hand, the low energy photons from the disk component are more sensitive to the
accretion rate rather than the inner radius. Thus the low energy photons from the disk and
the high energy ones from the power-law component have relatively less time delay as compared
to photons with energy close to $T_{in}$, which depend significantly on the inner radius
which is assumed to be delayed with respect to the accretion rate. Thus,  the turnover energy
depends sensitively on the flux ratio of the power-law and disk components characterized by
the parameter $N$. Since the ratio can be estimated from the time-averaged spectrum, there
is little leverage to tune the other parameters to get the observed lag turnover energy.

In Figure (\ref{betavar}), we show the rms and time-lag for variations for another important
parameter $\beta$ which is the exponent specifying the relation between the inner radius and the
accretion rate. It is interesting to note the sensitive behaviour of the time-lag for the next harmonic
to $\beta$ where small changes in $\beta$ lead to dramatic qualitative changes i.e. from hard to
soft lags. In Figure (\ref{powvar}), are shown the rms and time-lag variations for different photon indices.
The behaviour is similar to that observed from the Figure(\ref{Nvar}) since the ratio $\frac{F_{p}}{F_d}$
given by Eqn (\ref{eqndft1}) depends on $N$ and $p$.

\subsection{Comparison with observations}

\begin{figure}  
  \centering
   \includegraphics[width=0.3\textwidth,angle=-90]{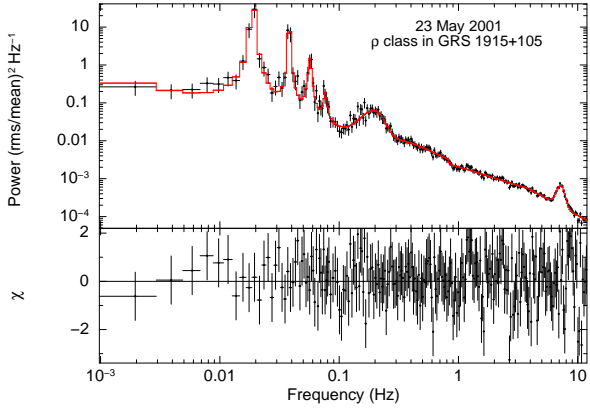}
\caption{Power density spectrum shown for the ``$\rho$'' variability class observed on 23 May, 2001 (OBS-Id: 60405-01-02-00), fitted with Lorentians and broken power-law along with residuals.} 
\label{powspec}    
\end{figure}

The `$\rho$' variability class shows a range of frequencies from
$\sim$ 10-20 mHz with the occasional presence of multiple
harmonics. Using RXTE/PCA archival data for the white noise-subtracted, rms-normalized power density
spectrum observed during `$\rho$' class on 23 May, 2001 (OBS-Id: 60405-01-02-00)  is shown
in the top panel of Figure \ref{powspec} fitted with multiple Lorentians and a
power-law with a break frequency at $\sim$ 0.1 Hz. Residuals of the
fit are shown in the lower panel.  From the fitting, the fundamental
and the next three harmonics are clearly detected with $>$ 3$\sigma$
significance at 19.5$^{+1.8}_{-1.2}$ mHz, 39.1$^{+2.5}_{-2.5}$ mHz,
58.6$^{+5.4}_{-4.7}$ mHz and 78.2$^{+6.5}_{-5.3}$ mHz
respectively. 

For the above observation, we obtain the time averaged photon spectrum from
{\tt standard2} data file of PCA. For data extraction, we use PCU2
only since it is reliably on throughout the  observation and has the  best
calibration accuracy. Using PCA responses and latest background
spectral file, the source spectrum in the energy range 3.0$-$25.0 keV
is fitted with a two component model : multi-colour disk blackbody ($\tt
diskbb$ in $\tt XSpec$) model and a simple power-law  ($\tt
powerlaw$ in $\tt XSpec$), along with a Gaussian component at $\sim$
6.4 keV and all components are modified by Galactic absorption ($\tt
TBabs$ in $\tt XSpec$). The fit gave a  $\chi2/dof = 40.41/40$ and the  best fit spectral parameters  are
listed in Table (\ref{tabledat}).

\begin{figure*}
 \centering
    \includegraphics[width=0.3\textwidth,angle=-90]{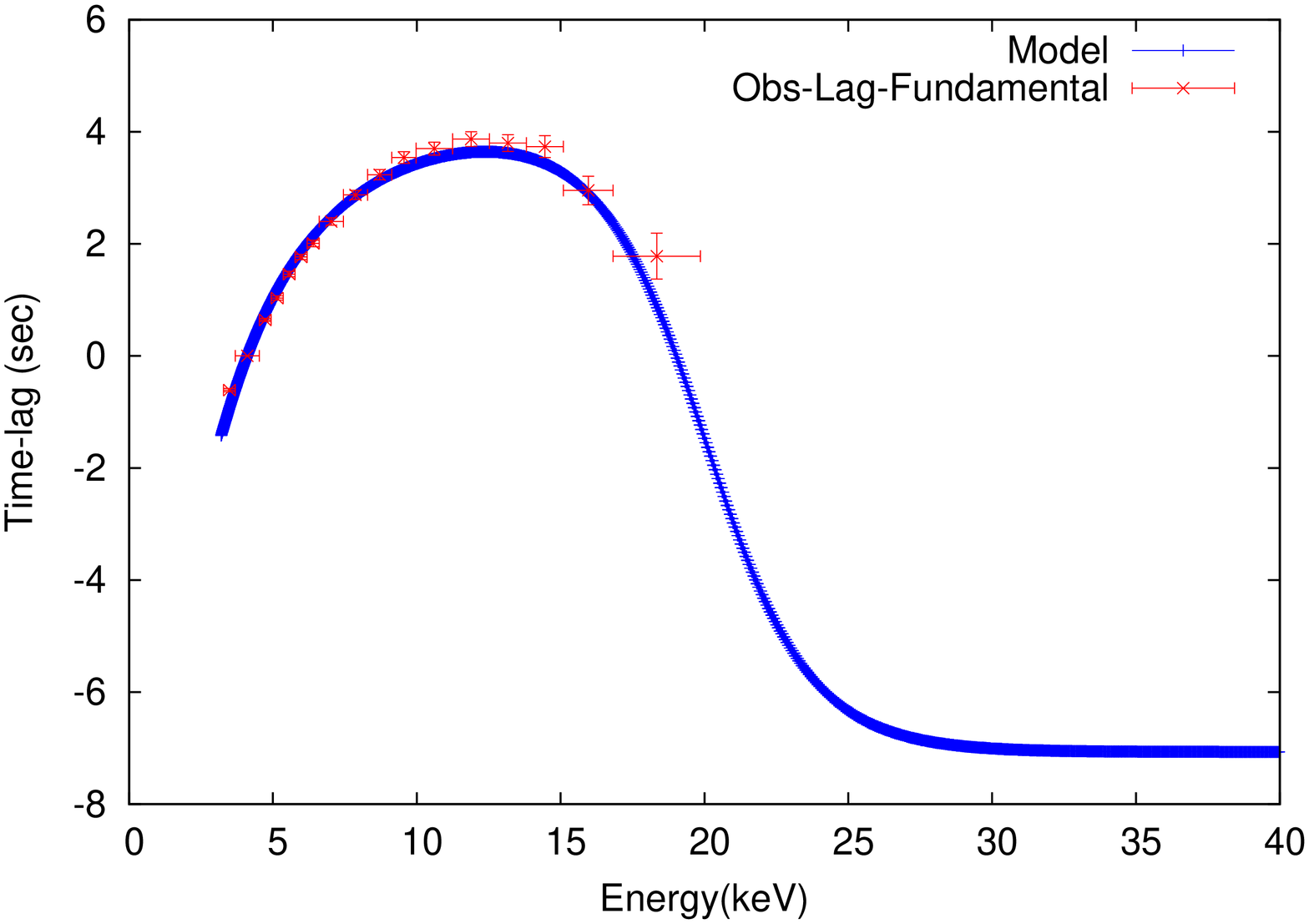}
    \includegraphics[width=0.3\textwidth,angle=-90]{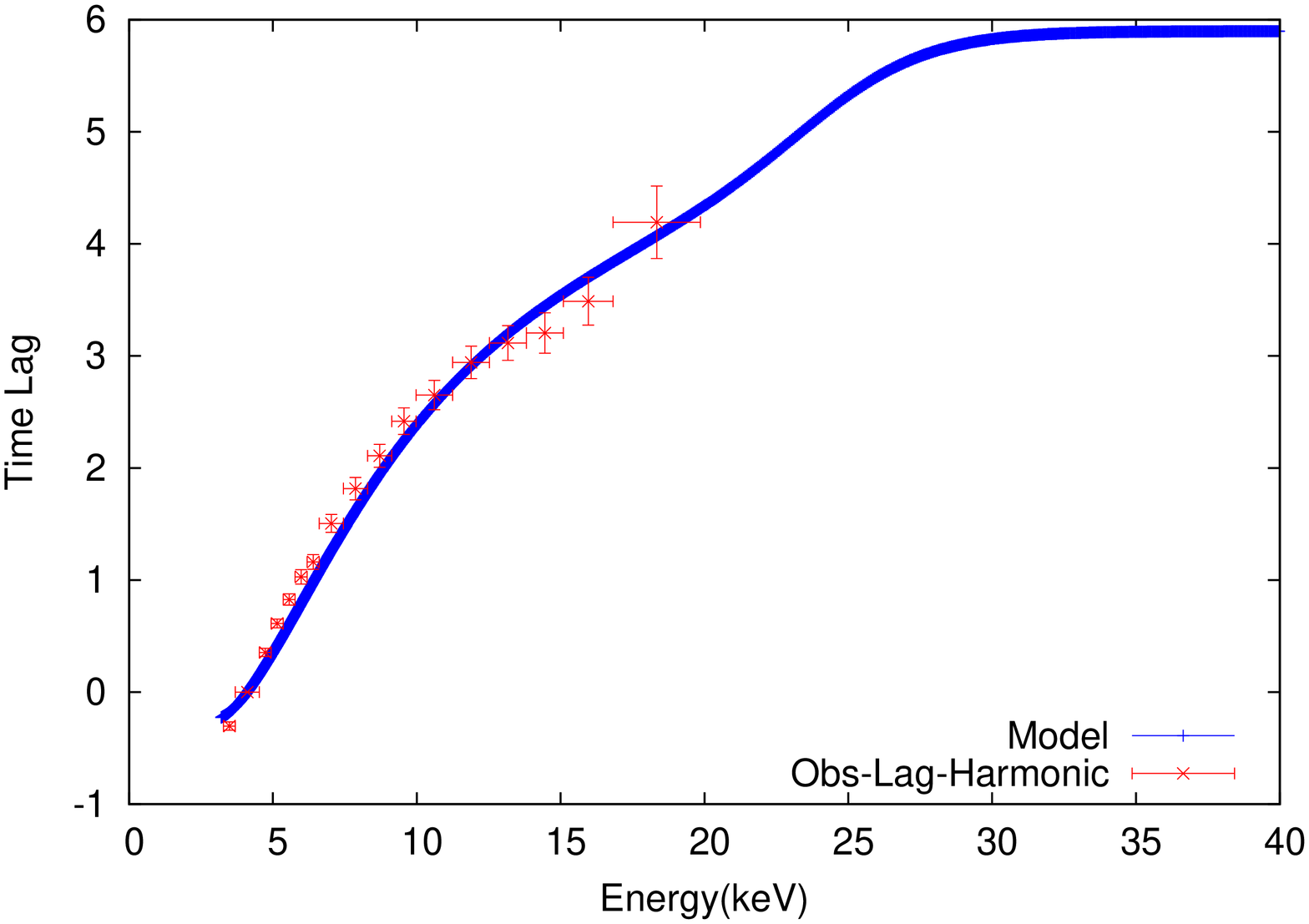}
          \includegraphics[width=0.3\textwidth,angle=-90]{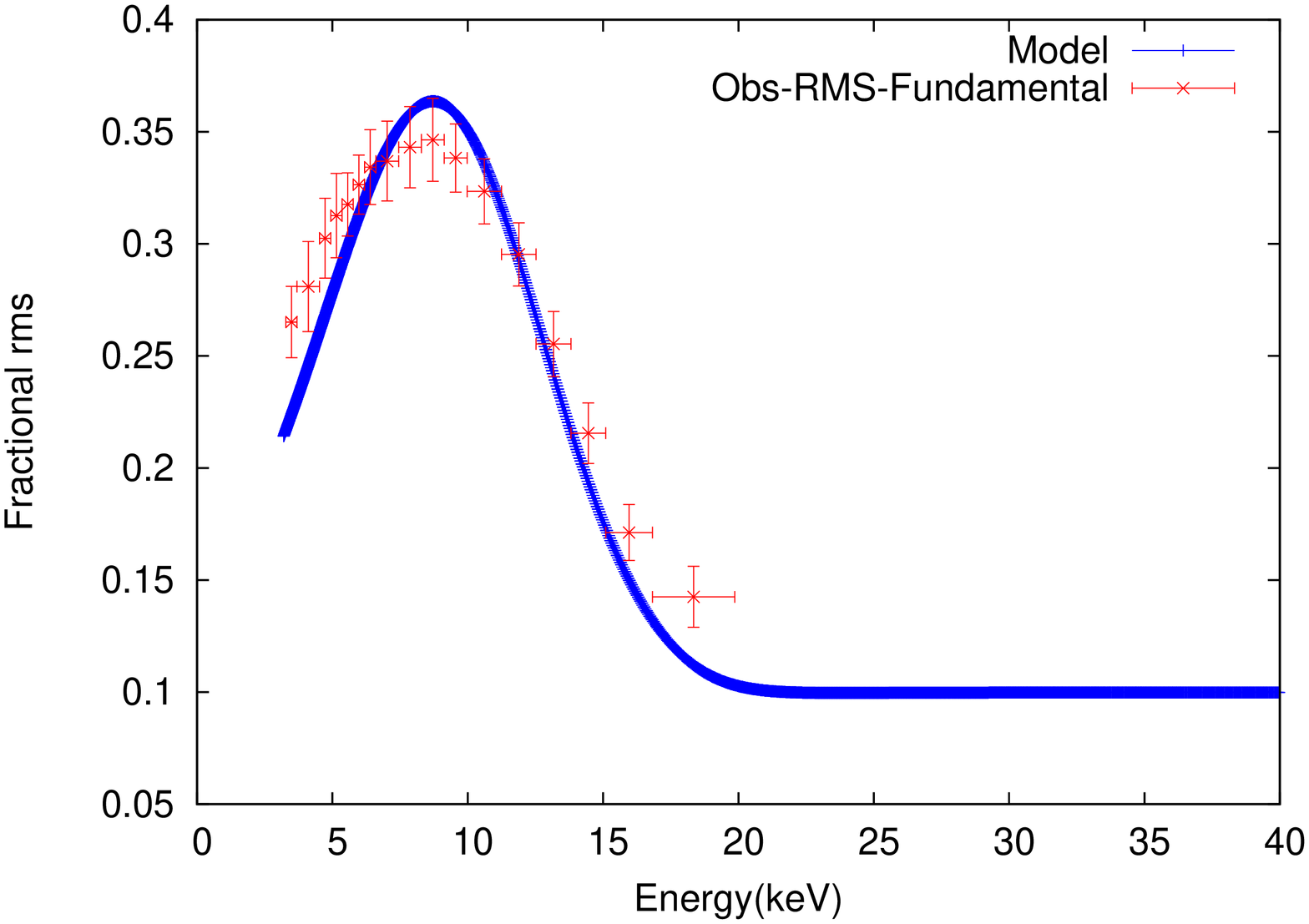}
          \includegraphics[width=0.3\textwidth,angle=-90]{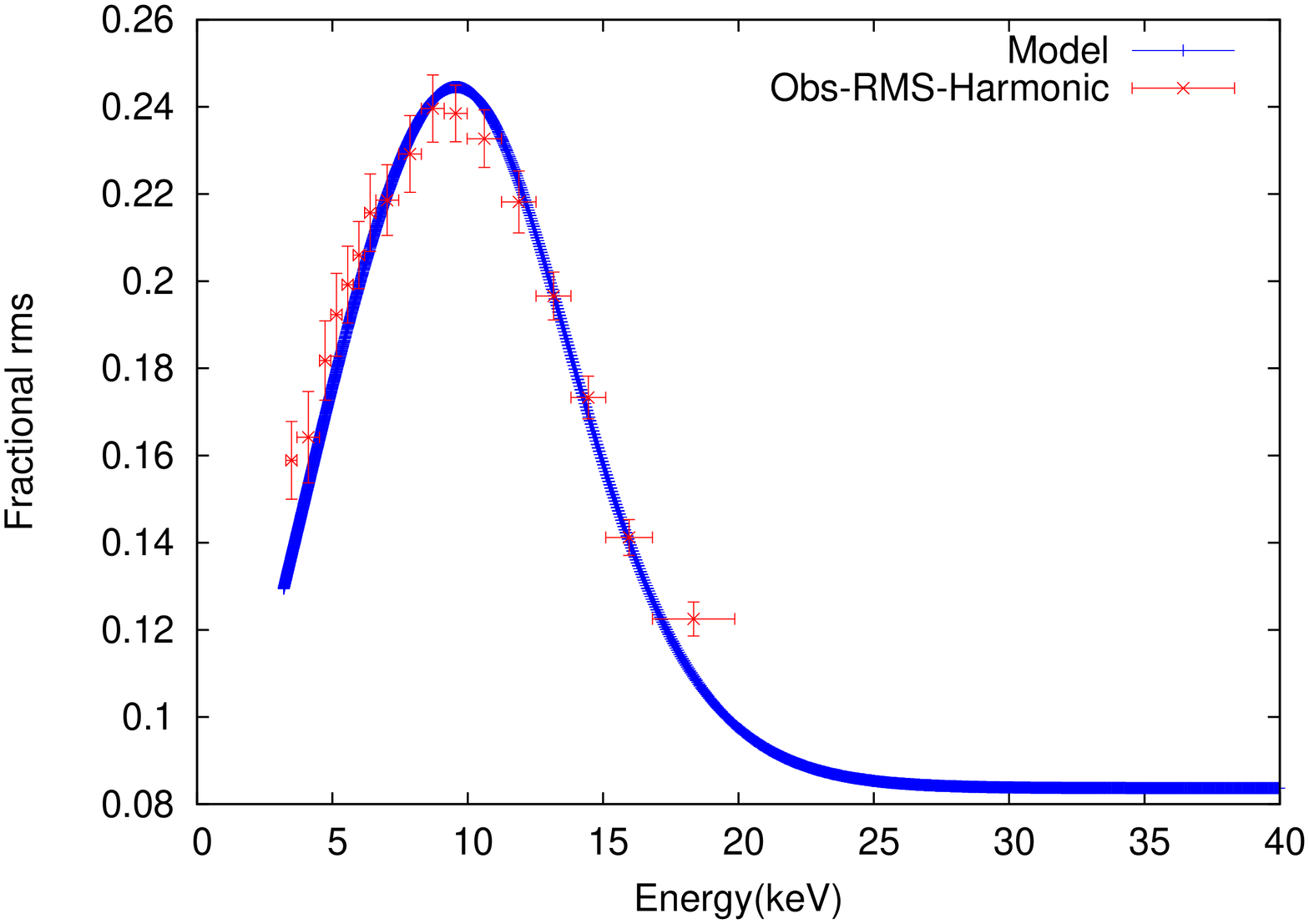}
  \caption{The observed and predicted energy dependent time-lag and rms at the fundamental (left panels) and for the
next harmonic (right panels) observed on 23 May, 2001. The parameters used for the model are listed in Table \ref{tabledat}. The four predicted curves
are determined by only four free parameters}
\label{datafig}
\end{figure*}

We calculate time-lag spectra using cross spectrum techniques
\citep{No99} in different energy bands in the range 3.20 $-$ 19.85 keV, where the reference band for lag calculation is chosen to be 3.69 $-$ 4.52 keV at
both fundamental and harmonic frequencies. The rms was calculated by computing the power spectrum for different energy bins.
Each power spectrum was fitted with a broken power-law and Lorentzians as shown in Fig. \ref{powspec}. Then by integrating the fitted components, the rms was calculated.  Table (\ref{tabledat}) shows the  values of the  
parameters used to model the observed lag and fractional rms. 
 As can be seen in Fig.(\ref{datafig}), the time-lag at fundamental (top left panel) shows a reversal around $\sim$ 12
keV while for the next harmonic (top right panel) the time-lag seems to rise at least up to $\sim$ 20 keV. However, the model-predicted spectrum fits the data well and the extension of the fitted
model till 40 keV shows saturation around 25 keV.
Because of poor coherence and poor statistics above 20 keV for RXTE/PCA, we are unable to verify the model-predicted saturation using actual data.

While a detailed analysis of the energy dependent rms and time-lag for different observations during the
$\rho$ class variability will be presented elsewhere (Mir et. al. to be submitted), here we have
used one observation as a typical  example. We note that the qualitative behaviour of the  energy dependence is 
fairly generic. Indeed, the
detailed phase resolved spectroscopy of different observations also show that the energy
dependence is similar despite changes in the frequency of the oscillations \citep{Ne11,Ne12}

\begin{table} 
\caption{Best fit  parameters obtained from time-averaged spectrum fitting and from the proposed model fitting of the energy dependent time-lag and rms for the 23 May 2001 observation of GRS 1915+105 during
the $\rho$ class}
\begin{center}

\begin{tabular} {c c c cc}
\hline  
 kT$_{in}$ &  N$_{d}$ & $\it p$ &$N_{p}$ & N  \\ 
(keV) & & & & \\
\hline
1.47$_{-0.01}^{+0.01}$ & 220$_{-11.14}^{+11.10}$ & 2.85$_{-0.02}^{+0.02}$ &34$_{-2.4}^{+2.3}$& 3.15$_{-.25}^{+.25}$   \\
\hline
$\beta$  & $\omega \tau_d$ & $\Gamma$ &$\delta \dot{M}$  \\ 
\hline
0.73 & 2.25  & 0.1 & 0.47 \\

\hline

\end{tabular}
\end{center}
Note:  kT$_{in}$ is the inner disk temperature (keV), ${\it p}$ is the photon power-law index, $N_{d}$ is the normalization of the
 disk blackbody component  $N_p$ is the power-law component normalization and $N$ is obtained from the spectral information. 
$N$ is related to the ratio between the normalizations of power-law and disk blackbody (Appendix B).
$\beta$ is the power-law variation of $R_{in}$ with $\dot{m}$ , $\omega t_d$ is the phase  difference, 
$\Gamma$ is the dependence of power-law normalization on $\dot{m}$ and $\delta \dot m$ is the amplitude of the
accretion rate variability.
\label{tabledat}
\end{table}

\section{Discussion and CONCLUSION}
GRS 1915+105 shows a fascinating clockwork in its $\rho$-class,
reflected in the highly periodic light curves which resemble closely
to the human cardiogram, thus is also called as heartbeat state. We
have studied the energy-dependent frequency resolved lags and
fractional rms of this variability and they show a unique
behaviour. With the lags of the order of few seconds, they cannot be
attributed  to light crossing time effects or due to Comptonization delays. The interesting feature
is that the lag at the fundamental frequency show a phase reversal
$\sim$ 10keV while at the next harmonic it keeps increasing with energy. The fractional rms amplitude
at the fundamental and the next harmonic are also non-monotonic with energy.

We show, that these qualitative features can be explained in a model
where the power-law component varies directly with the accretion rate
while the inner disk radius responds after a time delay (DROID). Apart from parameters
estimated from the time averaged spectrum, the energy dependent rms and time
lag for the fundamental as well as the next harmonic are specified by only four
free parameters and hence it is remarkable that such a simple picture can
even qualitatively reproduce the overall features.

It should be emphasized that the analysis of energy dependent time-lag and rms and
the model proposed to explain them are not different but complimentary to the
phase resolved spectroscopy done earlier by \citet{Ne11,Ne12}. Indeed as expected both
analysis find that the basic feature of the oscillation is that the inner disk
radius oscillates. The analysis done here provides another way of analyzing and
understanding the data. Moreover, for observations of other types of oscillations,
energy dependent rms and time-lag may be more easily estimated than performing
a complete phase resolved spectroscopy. 

A remarkable feature of this model is that the primary driver is a pure sinusoidal
oscillation in the accretion rate and the other harmonics can be attributed to
the non-linear relationship between the accretion rate and the inner disk radius
as well as that between the two and the emergent spectrum. As the accretion rate
increases the disk extends inwards, but this does not happen instantaneously.
The inner disk responds after a time-delay that is found for this observation to be $\sim 20$ secs, which
may be noted to be of the order of the viscous timescale.  Thus our results present
a scenario which has the potential to be developed into a physical interpretation based
on the nature of the  accretion disk instability.

The spectral components used in this model are simplistic. Instead of a power-law, the
high energy component should be correctly  modeled as a thermal Comptonization spectrum
where the Comptonized flux should depend on the accretion rate. This will lead
to variations in the spectral index which is considered in this work to be 
a constant. Moreover, there should also be a reflection component which
may play an important part in the QPO formation (e.g. \citep{SZ84}). The time-lag measured and modelled in this work
is of the order of few seconds. Unless
the reflector is significantly far away from the central object, any time-lag between the continuum and the reflection component (which is of the order sub-millisecs to few millisecs for a $\sim$ 10 $M_{\odot}$ black hole) 
 can be neglected for the analysis done here. However, the rms spectra obtained from AGN (e.g., MCG 6$-$30$-$15)
 do show significant dip \citep{ZY10} at the position of 6.4 keV Fe emission line which is solely due to reflection. Therefore, instead of a powerlaw the
combined continuum and reflection component needs to be considered, which will effect the
results primarily in the Iron line region \citep{Fa09}.
Such analysis would necessarily involve complex numerical computation which is presently out of scope of this work but can be considered as an extension of the present idea in future.
It is likely that the deviation of the data from the predicted model, especially for
the shape of the energy dependent rms curves, arises because of the simple
spectral model used. A more realistic model would warrant fitting the data statistically
and obtaining parameter values with confidence limits. It will be interesting to study
how the parameters of the model change for different oscillations of the $\rho$ class,
especially as a function of the frequency of the oscillations. Such a study will
shed light on structural changes that occur during these oscillations and will help
in obtaining a complete hydrodynamic understanding of the phenomenon.

Finally, the model maybe applicable to other QPOs observed in different systems
where the disc component varies significantly during the oscillation. Observations of
different black hole systems by the recently
launched Indian multi-wavelength satellite ASTROSAT is expected to  provide high quality event
tagged data  which would be ideally suited for such analysis.

\section{ACKNOWLEDGEMENTS} 
MHM is highly thankful to IUCAA, Pune for allowing the periodic visit to the
institute which has helped in carrying out this work and to UGC, New Delhi for granting the research fellowship.

\appendix

\section{Accretion rate induced spectral variability of a disk and a power-law components}

The accretion disk flux can be rewritten as
\begin{equation}
 F_D \propto  E^{-2/3} \dot{m}^{2/3} \int_{Z}^{\infty} \frac{y^{5/3}}{exp(y)-1}dy 
\end{equation}
where $Z \equiv E/kT_{in}$ and the relation $T_{in} \propto \dot{m}^{1/4}{r_{in}}^{-3/4}$ has been used. Its variation is
given by
\begin{eqnarray}
\delta F_D & = & (F_D)_{\dot{m}}\delta{\dot{m}} + (F_D)_{Z} \delta{Z} + \frac{1}{2}(F_D)_{\dot{m}\dot{m}}\delta{\dot{m}^2} \nonumber \\
& & +(F_D)_{\dot{m}Z}\delta{\dot{m}}\delta{Z}+\frac{1}{2}(F_D)_{ZZ} \delta{Z}^2 
\end{eqnarray}
Here and as has been used in the main text, we denote the normalized variation 
$\delta X = \Delta X/X$ and define 
\begin{equation}
(Z)_X = \frac{X}{Z}\frac{\partial Z}{\partial X}\;\;\;\; \hbox{and}\;\;\;\;  (Z)_{XY} = \frac{XY}{Z}\frac{\partial }{\partial X}\frac{\partial Z}{\partial Y}
\end{equation}
The derivatives can be computed to be $(F_D)_{\dot{m}} = 2/3$, $(F_D)_{\dot{m}\dot{m}} = -2/9$,
\begin{equation}
(F_D)_{Z} = f(Z) = \frac{Z^\frac{8}{3}}{({e^Z-1}){\int_Z^{\infty} \frac{y^\frac{5}{3}}{exp(y)-1}\,dy}}
\end{equation}
\begin{equation}
(F_D)_{ZZ} = f(Z)(\frac{5}{3} -k(Z)) = f(Z)(\frac{5}{3} -\frac{ze^Z}{e^Z-1})
\end{equation}
and $(F_D)_{\dot{m}Z} = 2f(Z)/3$.

Since Z depends on $T_{in}$ which in turn depends on $\dot{m}$ \& $r_{in}$, as $Z=\alpha\dot{m}^{-1/4}r_{in}^{3/4}$,
we have:
\begin{equation}
\delta Z = \frac{3}{4}\delta r_{{in}}-\frac{1}{4}\delta {\dot{m}}-\frac{3}{32} (\delta r_{{in}})^2+ \frac{5}{32} (\delta \dot{m})^2-\frac{3}{16}\delta \dot{m} \delta r_{{in}}
\end{equation}

The power-law component is assumed to depend on the accretion rate as $F_p \propto \dot{m}^\Gamma E^{-P}$ and hence
\begin{equation}
\delta F_p = \Gamma \delta \dot{m} +\frac{\Gamma(\Gamma-1)}{2} (\delta \dot{m})^2
\end{equation}
The variation in the total flux $F_t = F_D+F_p$ is
\begin{equation}
\delta F_t = \frac{F_D}{F_t} \delta F_D  + \frac{F_p}{F_t} \delta F_p
\end{equation}

Next assuming that the inner disk radius follows the accretion rate with a
time delay $r_{in}(t) \propto \dot{m}^{-\beta}(t-\tau_d)$
and that the accretion rate undergoes a pure sinusoidal variation at an
angular frequency $\omega$ i.e. $\delta \dot m = |\delta\dot{m}|e^{i \omega t}$
the inner disk variation is given by
\begin{equation}
\delta r_{in} =  -\beta|\delta\dot{m}| e^{i\omega(t-\tau_d)} +\frac{\beta (\beta+1)}{2}|\delta\dot{m}|^2 e^{2i\omega(t-\tau_d)} 
\end{equation}

Finally collecting terms proportional to $e^{i\omega t}$ and $e^{2i\omega t}$, we get
for the fundamental oscillation,
\begin{equation}
\delta F_t^f  = \frac{|\delta \dot{m}|}{1+\frac{F_p}{F_d}}\left(Ae^{i\omega t} + Be^{i\omega (t-\tau_d)}\right)
\end{equation}
and for the next harmonic:
\begin{equation}
\delta F_t^h=\frac{|\delta\dot{m}|^2}{1+\frac{F_p}{F_d}}\left(A^\prime e^{2i\omega t}+B^\prime e^{2i\omega(t-\tau_d)}+C^\prime e^{2i\omega(t-\tau_d/2)}\right) 
\end{equation}
where
\begin{align*} 
&A=\frac{2}{3}+\frac{f(z)}{4}+\Gamma\frac{F_p}{F_d}\\ 
&B=\frac{3}{4}\beta f(z)\\
&A^\prime=-\frac{1}{9}+\frac{f(z)}{8}\left(\frac{k(z)}{4} - \frac{1}{3}\right)+ \frac{1}{2}\Gamma(\Gamma-1)\frac{F_p}{F_d}\\
&B^\prime=\frac{3}{8}f(z)\left(\beta^2\left(\frac{3}{4}k(z)-1\right)-\beta(\beta+1)\right)\\
&C^\prime=\frac{3}{16}\beta k(z)f(z)
\end{align*}

\section{Estimating the power-law to disk flux ratio from time-averaged spectrum} 

In terms of $Z= E/kT_{in}$, the ratio of the power-law to disk flux can be 
written as
\begin{equation}
\frac{F_p}{F_d} = N G(Z,p) = N Z^{-(p+2)} f(Z) (e^Z-1)
\label{eqndft1}
\end{equation}
From {\it Xspec} model fitting of the spectrum we get the normalization of the
power-law $F_p = N_p E^{-P}$ and that of the disk emission,
\begin{equation}
N_{d}=\left(\frac{(r_{in}/kms)}{(D/10kpc)}\right)^2cos(\theta)
\end{equation}
By carefully comparing the flux equations, substituting $Z$ for $E$ and 
converting  the energy unit from keV to ergs, one gets
\begin{equation}
 N= \frac{3}{16\pi}\frac{N_p}{N_d}(kT_{in})^{-(p+2)}hc^2 \left(\frac{10^5}{10kpc}\right)^{-2} \left(\frac{ergs}{keV}\right)^{-3}
\end{equation}
or numerically
\begin{equation}
N \approx 1.316\times 10^2 \times \frac{N_p}{N_d}\left(\frac{kT_{in}}{1 keV}\right)^{-(p+2)}
\end{equation}

\bsp	
\label{lastpage}
\end{document}